\documentclass[prl,aps,floats, showpacs]{revtex4}
\usepackage{epsfig}
\usepackage[all]{xy}
\usepackage{graphicx}
\usepackage{latexsym}

\def\T0{{\bar{ T_0}}}

\newcommand{\be}{\begin{equation}}
\newcommand{\ee}{\end{equation}}
\newcommand{\bea}{\begin{eqnarray}}
\newcommand{\eea}{\end{eqnarray}}
\newcommand{\bml}{\begin{mathletters}}
\newcommand{\eml}{\end{mathletters}}

\begin{document}

\preprint{LPT-03-100}
%\draft
\tighten
\title{Gauss-Bonnet gravity renders negative tension braneworlds unstable} 
\author{Christos Charmousis and Jean-Fran\c{c}ois Dufaux}
\affiliation{LPT, Universit\'e de Paris-Sud, B\^at. 210, 91405 Orsay CEDEX, France}
%\date{\today}
\setlength{\footnotesep}{0.5\footnotesep}
%\maketitle

\begin{abstract}
We show that the addition of the Gauss-Bonnet
term to Einstein gravity induces a tachyon mode in the 
spin 2 fluctuations of the Randall-Sundrum I model. 
We demonstrate that this instability is generically related to the presence 
of a flat negative tension brane, of co-dimension one, embedded in an anti-de
Sitter background. In particular its presence is independent of $Z_2$-symmetry or 
compactness of the extra dimension. The gravitational tachyon mode
, persists for arbitrarily small but non vanishing 
Gauss-Bonnet coupling. It is a bound state localised on the negative
tension brane, much like the graviton zero-mode is localised on a
positive tension one. We discuss the possible resolution of this
instability by the inclusion of induced gravity terms on the branes or
by an effective four-dimensional cosmological constant.
\end{abstract} 
\pacs{04.50.+h, 11.25.Db, 11.25.Mj}

\maketitle

%\newpage

%\vspace*{0.5 cm}

Negative tension branes appear quite naturally as endpoints of spacetime 
in higher than 4 dimensions. For example they may appear in 
non-oriented string theories as orientifold planes \cite{sagnotti}, 
\cite{polc}, or in orbifold compactifications of braneworld models.
In the Randall-Sundrum \cite{RS1} model (RS1), their embedding in anti-de
Sitter spacetime provides an elegant resolution of the hierarchy
problem, once the interbrane distance is fixed \cite{GW}. Interestingly, 
any realisation of such a geometrical hierarchy in string theory seems
to also require the presence of negative tension objects \cite{Pol}, 
orientifolds.

Higher order curvature
corrections also appear in string theory as $\alpha'$
corrections in low energy effective actions \cite{gross}. 
The Gauss-Bonnet term in particular, appears (for
example) in heterotic string theory rendering the low energy effective 
action ghostfree \cite{Zwieb} around a flat background. 
More importantly, in a non-perturbative approach, one obtains
such a curvature invariant
 from purely geometric considerations in higher than 4
dimensions. Indeed the Gauss-Bonnet combination is the
unique higher order curvature term, in five or six dimensions, leading to a 
classical gravity theory which satisfies the physical assumptions of
General Relativity in four dimensions \cite{Love}. In particular, the  
resulting field equations involve only up to second order derivatives of 
the metric. This is an essential assumption when treating 
backgrounds with boundaries, and it renders spacetime perturbations wavelike. 

In this letter we consider braneworld models involving a negative
tension brane, with four-dimensional Poincar\'e invariance, embedded in 
five-dimensional anti-de Sitter (adS) background. We will show that the
Gauss-Bonnet term induces generically at least one tachyon mode in
the spin 2 fluctuations of such models.

Consider a model with two branes in Einstein-Gauss-Bonnet (EGB) gravity :
\be
\label{Sbulk}
S=\frac{M^3}{2} \int d^5x \sqrt{-g} \,
\left(-2\Lambda+R+\alpha\,
\left[R^2-4 R_{ab}R^{ab}+ R_{abcd}R^{abcd}\right] \right)
-\int d^4x\, \sqrt{-\gamma_1} \,\, T_1
-\int d^4x \,\sqrt{-\gamma_2} \,\, T_2
\ee
We will take $\alpha\geq 0$ in the following
as in the case of string slope expansion \cite{gross}. 
The solution with a smooth (Einstein) $\alpha \rightarrow 0$ limit 
and four-dimensional Poincare invariance reads,
\be
\label{horo}
ds^2=e^{2 A(y)}\,\eta_{\mu \nu}\,dx^{\mu}dx^{\nu}+dy^2
\ee
where 
\be
\label{warp}
A(y)=-ky \hspace*{0.5cm} ; \hspace*{0.5cm}
k=
\sqrt{\frac{1}{4\alpha}\left(1-\sqrt{1+\frac{4}{3}\alpha\Lambda}\right)}
\ee 
The solution is defined for $4\alpha k^2 \leq 1$. The two flat branes have 
tension $T_1$ and $T_2$, and are located at $y=0$ and $y=y_c$ respectively.
We also assume $Z_2$-symmetry accross the corresponding orbifold fixed points. 
The junction conditions \cite{stephen} then require,
\be
\label{tens}
\frac{T_1}{M^3}=-\,\frac{T_2}{M^3}=2k(3-4\alpha k^2)
\ee
where $T_1>0$ and $T_2<0$.
This setup is just the generalisation, in Gauss-Bonnet gravity, of
the RS 1 model \cite{RS1} where typically,
$ e^{k y_c} \sim 10^{15}$. 

Let us now consider the spin 2 fluctuations around this background.
For these it is sufficient to 
consider a linear perturbation in Gaussian normal gauge,
\be
\label{perturb}
ds^2=e^{2A(y)}\,\left(\eta_{\mu \nu}+h_{\mu \nu}^{(m)}(x)\,\psi_m(y)
\right) + dy^2
\ee
with $h_{\mu \nu}^{(m)}(x)$ transverse and tracefree. 
The metric fluctuations verify, 
$\Box_{(4)} h_{\mu \nu}^{(m)}(x) = m^2 h_{\mu \nu}^{(m)}(x)$, 
leading to plane wave separation of variables, 
$h_{\mu\nu}^{(m)}(x)=\epsilon_{\mu\nu} e^{i p_\lambda x^{\lambda}}$ 
with 4-dimensional momenta $p^2=-m^2$,  where $\epsilon_{\mu\nu}$ is the
constant polarisation tensor. We do not consider here the scalar mode
describing the interbrane distance \cite{radion}, the radion, which will
not be relevant for our purpose{\footnote{a direct calculation in the lines of
\cite{radion} shows that the radion mode remains a
massless scalar in Gauss-Bonnet gravity}}. The perturbation equation in
the bulk reads, 
\be
\label{wave}
-\left(p(y)\psi_m'(y)\right)' = m^2 \, w(y) \, \psi_m 
\ee
where a prime denotes derivative with respect to $y$ and 
\be
\label{pwg}
p(y)=e^{4A}(1-4\alpha A'^2), \quad w(y)=e^{2A}(1-4\alpha A'^2-4\alpha A'')
\ee
Because of the Gauss-Bonnet term, $p$ and $w$ now involve
derivatives in the background field $A(y)$, which are not continuous in
the presence of branes. For example, $w=\hat{w}+[-4\alpha
e^{2A} A']\,\delta(y-y_i)$, where $\hat{w}$ stands for the continuous
part of $w$, and $[\;]$ denotes the jump across the brane location
$y=y_i$ (see also for example \cite{neu}). 
For the two brane system we consider here, we have in the bulk ($0<y<y_c$),
\be
\label{pw}
p(y)=(1-4\alpha k^2)\,e^{-4ky}, \quad \hat{w}(y)=(1-4\alpha k^2)\,e^{-2ky}
\ee
while the junction conditions resulting from (\ref{wave}) read:
\bea
\label{bound1}
\psi_m'(0^+)=-\frac{4\alpha k}{1-4\alpha k^2}\,m^2\,\psi_m(0)
\\ \label{bound2} 
\psi_m'(y_c^-)=-\frac{4\alpha k}{1-4\alpha k^2}\,m^2\,
e^{2 k y_c}\,\psi_m(y_c)
\eea
For $\alpha=0$ (or $m=0$), these are usual Neumann boundary
conditions. In Gauss-Bonnet gravity however, we have mixed boundary
conditions involving the energies of the modes themselves. 
As a result, their norm $||\psi_m||$ has to include suitable boundary 
terms (see also \cite{us}),
\be
\label{norm}
0 \leq \,\int_0^{y_c}dy\,p\,\psi'^2_{m} =
m^2\,\left[\int_0^{y_c}dy\,\hat{w}\,\psi_m^2  +
4\alpha k \left( \psi_m^2(0)-e^{-2k
y_c} \psi_m^2(y_c)\right)\right] = m^2\, ||\psi_m||^2
\ee
where the first equality results from multiplying (\ref{wave}) by
$\psi_m$, integrating by parts on $y$ and using
(\ref{bound1})-(\ref{bound2}). As usual, $p$ and $\hat{w}$ (\ref{pw})
are positive definite from 
(\ref{warp}). Therefore, for $\alpha=0$, (\ref{norm}) implies that $m^2$
has to be non negative. However, this standard positivity argument 
breaks down if any of the boundary terms in (\ref{norm}) is
negative as in the case of negative tension
branes. Tachyonic states may then appear in the spectrum. Note also that
the norm is negative in that case. In this sense any tachyon in the bulk
perturbations can be also interpreted as a ghost at the level of the
4-dimensional effective action. Indeed expanding
(\ref{Sbulk}) to second order we obtain,
\be
\label{effe}
S_{eff} \sim -\frac{M^3}{4} ||\psi_m||^2 \int d^4 x \; (\nabla h_{\mu\nu})^2 + 
m^2 \, h_{\mu \nu} h^{\mu \nu}.
\ee
For $m^2<0$ in (\ref{norm}) there is always a wrong relative sign
between the kinetic and mass terms which signals classical instability. 

\begin{figure}[htbp]
\begin{center}
\begin{tabular}{cc}
\includegraphics[angle=-90,width=.50\textwidth,bb=68 68 543 722]{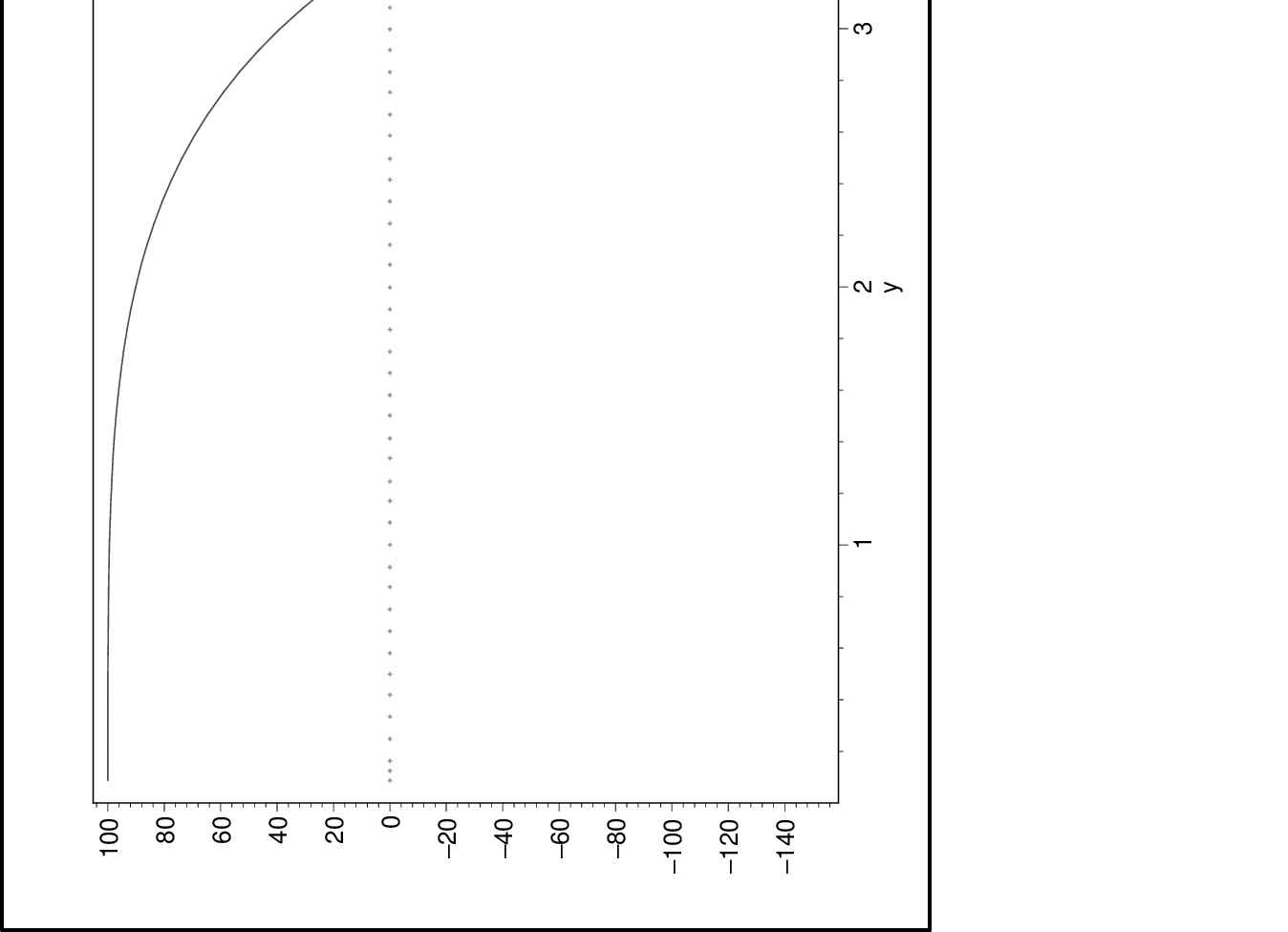}&
\includegraphics[angle=-90,width=.50\textwidth,bb=68 68 543 722]{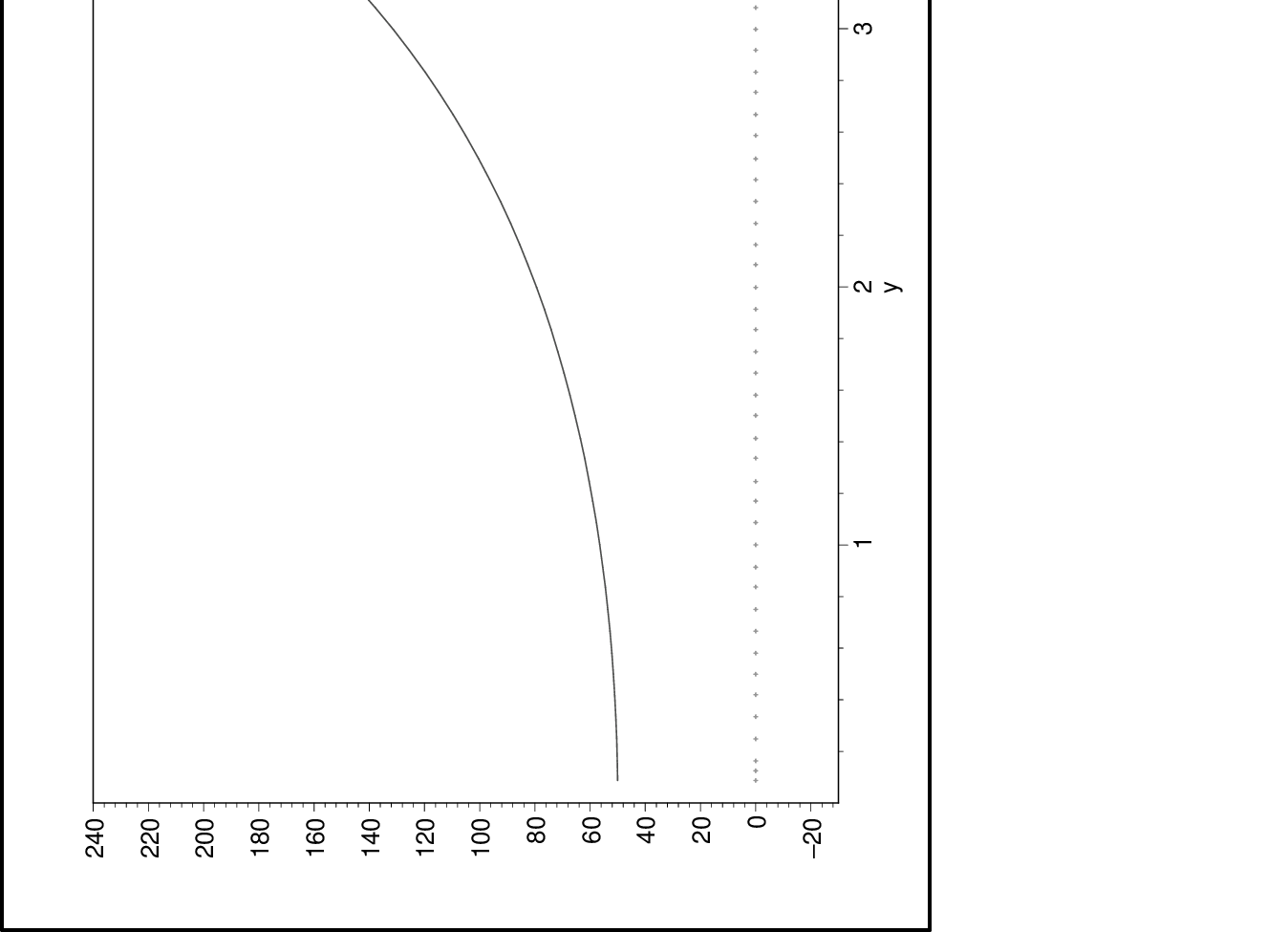}
\end{tabular}
\caption{$f$ (line) and $g$ (dots) as functions of $\chi x$ for
$\zeta=0.5$ (left) and $\zeta=0$ (right) ; $\chi=100$ here.} 
\label{fig31}
\end{center}
\end{figure}

We now show that such a tachyon state is indeed present for the
two brane system. 
Any solution of (\ref{wave}) with $m^2=-\mu^2 < 0$ is of the form,
\be
\label{solmu}
\psi_{\mu}(y)=e^{2k y}\,
\left( A_{\mu} \, I_2\left(\frac{\mu}{k}e^{k y}\right) 
+ B_{\mu} \, K_2\left(\frac{\mu}{k}e^{k y}\right) \right)
\ee
with $A_{\mu}$ and $B_{\mu}$ real constants. 
A linear perturbation containing such an  eigenmode 
will grow exponentially with time and will signal the classical
instability of the background. A tachyon mode of mass $\mu$ exists if
on imposing (\ref{bound1})-(\ref{bound2}) on (\ref{solmu}), 
$x=\frac{\mu}{k}>0\,\, ,$ is solution of (for $\mu \neq 0$),
\be
\label{cond}
f(x)=g(x)
\ee
with
\bea
\label{f}
f(x)=\bigg(I_1(\chi x)-\zeta \chi x I_2(\chi x)\bigg)\,\,
\bigg(K_1(x)+\zeta x K_2(x)\bigg)  \\
g(x)=\bigg(I_1(x)-\zeta x I_2(x)\bigg)\,\,
\bigg(K_1(\chi x)+\zeta \chi x K_2(\chi x)\bigg) \nonumber
\eea
We have introduced two numerical parameters: 
\be
\label{param}
\zeta := \frac{4\alpha k^2}{1-4\alpha k^2} \hspace*{1 cm}
\mathrm{ and } \hspace*{1 cm} \chi:=e^{k y_c} > 1 
\ee
The parameter $\chi$ describes the interbrane distance and is typically
a large number. Parameter $\zeta$ describes deviation from
General Relativity and we have $0 \leq \zeta 
< +\infty$.
For $\alpha\Lambda$ small, which essentially means dominant Einstein
gravity, $\zeta$ is also small, with $\zeta=0$ for
$\alpha=0$. Alternatively $\zeta$ goes to infinity for $\alpha \Lambda=-4/3$.
Apart from these limiting cases, it is easy to show that (\ref{cond})
always admits a finite non-vanishing solution for $x$. The functions 
$f(x)$ and $g(x)$ are shown in Fig.~\ref{fig31}. 

For $x\rightarrow 0$,
\be
\label{f0}
f(x) \rightarrow (\zeta+\frac{1}{2})\,\chi, \qquad
g(x) \rightarrow \frac{(\zeta+\frac{1}{2})}{\chi}
\ee
whereas as $x\rightarrow +\infty$,
\be
\label{finf}
f(x) \sim -\frac{1}{2}\,\zeta^2\,\sqrt{\chi}\,x\,e^{(\chi-1)x}, \qquad
g(x) \sim -\frac{1}{2}\,\zeta^2\,\sqrt{\chi}\,x\,e^{-(\chi-1)x}
\ee
Therefore 
$f(x) > g(x) > 0$ for $x\simeq 0$, while for $x\rightarrow +\infty$
$g(x)\rightarrow 0$ and $f(x) \rightarrow -\infty$ (note however that
when $\zeta=0$, $f(x) \rightarrow +\infty$ as $x\rightarrow +\infty$).
Since $f(x)$ and $g(x)$ are continuous on $]0 \,\, +\infty[$,
they have to intersect at least once at finite $x$, corresponding to the
tachyon mass. The instability
persists for arbitrarily small but non vanishing coupling $\alpha$.

\begin{figure}[htbp]
\begin{center}
\includegraphics[angle=-90,width=0.45\textwidth,bb=68 68 543 722]{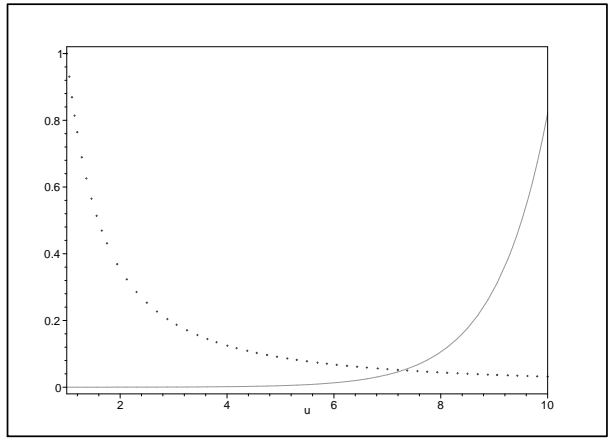}
\caption{Tachyon mode $\Phi_{\mu}(z)$ (line) localised on the TeV brane
and zero mode $\Phi_0(z)$ (dotted) localised on the Planck brane, for
the two branes system ($\chi=10$ and $\zeta=0.1$)}
\label{fig32}
\end{center}
\end{figure}

In order to illustrate the qualitative properties of the tachyon mode,
we go now to the analog quantum mechanical picture.
Consider 
$|z|=\frac{1}{k}(e^{k|y|}-1)$, 
with the branes now located at $z=0$ and $z=z_c$, and define 
$\Phi(z)=e^{-3k|y|/2}\,\psi(y)$. Then $\Phi(z)$ obeys the 
Schr$\ddot{o}$dinger equation,
\be
\label{schro} 
-\frac{d^2 \Phi}{dz^2} + V(z)\, \Phi = m^2 \, 
(1 + \frac{2\zeta}{k}\;\delta(z) 
- \frac{2\zeta\chi}{k}\;\delta(z-z_c)) \, \Phi
\ee
Note the presence of the 
mass-dependent distributional terms in the
right-hand-side. 
For the bound state solutions $m^2=-\mu^2\leq 0$, the effective
potential reads,
\be
\label{pot}
V^{\mu}_{\mathrm{eff}}=\frac{15 k^2}{4(k|z|+1)^2} - 
\bigg(3k - 2\;\frac{\zeta}{k}\;\mu^2 \bigg)\, \delta(z) 
+ \bigg(3 \; \frac{k}{\chi} - 2\;\frac{\chi}{k}\;\zeta\;\mu^2
\bigg)\, \delta(z-z_c) 
\ee
where we have included the energy-dependent distributional terms 
for the sake of illustration.  
Much may be understood from the distributional contributions to the potential 
(\ref{pot}), which are either attractive if their overall 
coefficient is negative, or 
repulsive if it is positive. The contribution centered on the Planck brane at 
$z=0$ is attractive for $\mu=0$, and it allows for the usual
normalisable bound state zero mode.
%{\footnote{This is to be contrasted with the 
%contribution centered on the $TeV$ brane at $z=z_c$, which is repulsive 
%for $\mu=0$ (or any $m^2>0$).}} 
However, we see that the contribution at $z=z_c$ may 
now be attractive for $\mu \neq 0$. Namely the negative tension brane
can support a new normalisable tachyonic bound state if $\zeta \neq 0$. 
The wave function of the two bound states is shown in Fig.~\ref{fig32}. 

We may approximate the mass and the wave function of the 
tachyon $\Phi_\mu$ by solving (\ref{schro}) in the vicinity
of the negative tension brane at $z=z_c$. Then, for $\chi>>1$, 
the bulk potential (\ref{pot}) is approximately constant
and we can neglect the boundary
term at $z=0$. To constant order in $\zeta$, the tachyon
mass is, 
\be
\label{jihad}
\mu \simeq {k\over \chi \zeta}(1+3\zeta)
\ee
The distributional term at $z=z_c$ in (\ref{pot})
is then indeed attractive. The corresponding wave function is localised on the
negative tension brane, much like the zero mode is bound on the positive
tension one. If one is to stabilize the hierarchy between the $TeV$ scale 
and the four-dimensional Planck mass $M_{\mathrm{Pl}}$ as in \cite{RS1}, 
then we have to live on the negative tension brane where the physical
masses are measured as $m_c=\chi m$ and $k \sim TeV$.
The approximation (\ref{jihad}) for the tachyon mass $\mu$ holds 
in particular in this case, and it gives $\mu_c \sim TeV$ if $\zeta \sim 1$, 
whereas $\mu_c \sim M_{\mathrm{Pl}}$ if $\zeta$ is fine-tuned to zero at the 
$10^{-15}$ level. Generically, 
the smaller the coupling $\alpha$, the larger the tachyon mass
and the more localised its wave function: {\it Even when the Gauss-Bonnet
coupling $\alpha$ is small, the Gauss-Bonnet term in the action 
does not act as a perturbative correction}. In particular, the limit $\alpha
\rightarrow 0$ is discontinuous for the tachyon mass, 
namely we have then $\mu\rightarrow\infty$, whereas there is no tachyon
when $\alpha=0$ exactly.

The instability of negative tension branes is independent of $Z_2$
symmetry or compactness of the extra dimension.
Consider a single $Z_2$-symmetric {\it positive} tension brane 
in an infinite extra dimension, ie. the second Randall-Sundrum model \cite{RS2}, in
EGB gravity. Any tachyon mode would still be 
given by (\ref{solmu}), but the boundary condition at infinity now requires 
$A_{\mu}=0$. 
Imposing the junction condition (\ref{bound1}) at $y=0$, 
we end up with the second factor of $f(x)$ in (\ref{f})
vanishing, which 
has no solution. This model has therefore stable spin 2
fluctuations. Accordingly, sending the negative tension brane to
infinity in (\ref{jihad}), ie. $\chi\rightarrow \infty$, sets $\mu=0$.
On the contrary if we choose to keep the {\it negative} tension brane, 
we have $B_\mu=0$ at infinity 
and we loose the zero mode keeping the tachyonic
one. Another illustration is provided by the  
Gauss-Bonnet version of the Lykken-Randall model 
\cite{LR}, where the tension of the second brane may be either positive or 
negative. In this setup we keep $Z_2$-symmetry across a first brane at $y=0$, 
but break it across a second brane at $y=y_c$. For $|y| < y_c$, the bulk
cosmological constant  is still denoted as $\Lambda$, and the background
solution is still given by (\ref{warp}). However, one 
introduces now another cosmological constant $\tilde{\Lambda}$ in the
bulk for $|y| > y_c$, with corresponding warp factor $\tilde{k}$ in the
background solution. Then, the tension of the
first brane is still given by $T_1$ in (\ref{tens}), while for the second
brane it reads, $T_2=M^3(\tilde{k}-k)\,\left[ 3-4\alpha k^2-4\alpha\tilde{k}^2 \right]$
which is positive or negative, according to the sign of $(\tilde{k}-k)$.
The candidate tachyon mode, is given seperately in 
two regions (\ref{solmu}), and as before we impose (\ref{cond}) with 
similar asymptotic behaviour, except that we now have,
\be
f(x) \sim
\frac{1}{2}\,\zeta^2\,\sqrt{\chi}\,x\,e^{(\chi-1)x}\,(\frac{\tilde{k}}{k}-1)
\quad \; \mbox{ for } \; x\rightarrow +\infty 
\ee  
Hence, when both branes have positive tension, that is when
$\tilde{k} > k$, $f(x)\rightarrow +\infty$ for $x\rightarrow +\infty$,
rather than minus infinity (\ref{finf}), and there is no tachyon. 
In the contrary if one of the branes has  
negative tension, we have tachyonic instability. 
It is therefore clear that in
anti-de Sitter background, any inclusion of a flat negative
tension brane destabilises the system in EGB gravity. 
    
The presence of induced gravity terms on the branes can modify the
above conclusions (see also \cite{dav}).
%\footnote{Consequences of brane induced gravity added to
%the RS1 model have been studied for instance in \cite{dav}. A
%cosmological model with one brane, with both induced
%gravity and Gauss-Bonnet corrections, has been considered in \cite{roy}. 
%}. 
Indeed, consider the action:
\be
\label{St}
S_{t}=S+\frac{M^3}{2}\bigg( \beta_1\,\,\int_{y=0}d^4x\, \sqrt{-\gamma}
\,\, R^{(4)}[\gamma] 
+ \beta_2 \,\, \int_{y=y_c}d^4x \,\sqrt{-\gamma} \,\, R^{(4)}[\gamma]\bigg)
\ee
Only the junction conditions for the massive modes are
modified. We can
follow the same analysis as above defining,
\be
\label{fbis}
\zeta_1 = \frac{4\alpha k^2+\beta_1 k/2}{1-4\alpha k^2} 
\,\,\, \mbox{ and } \,\,\, \zeta_2 =
\frac{4\alpha k^2-\beta_2 k/2}{1-4\alpha k^2} 
\nonumber
\ee
For simplicity we neglect terms of order $\frac{1}{\chi^2}$.
Requiring the effective four-dimenssional Planck mass squared to be
positive then gives $\zeta_1 \geq -1/2$. Furthermore, the presence of brane
induced gravity terms on the negative tension 
brane may lead to the radion field being a ghost 
(see for instance \cite{csaki}). This also signals instability, although
at the quantum level this time.
Generalising the method used in \cite{pilo} to the action (\ref{St}), we find
that for this not to be the case, the parameters have to satisfy 
$(1-4\alpha k^2)/(1+2 \zeta_2) \geq 0$. Taking these constraints into account, 
the range of parameters for which the gravitational tachyon mode may be 
avoided reduces to
\footnote{We thank Y.Shtanov for pointing out a
previous mistake in (\ref{range}), which was in contradiction with a
result in \cite{yuri}.}
\be
\label{range}
\zeta_1 \geq 0 \mbox{ and } -\frac{1}{2}\leq \zeta_2 \leq 0 \\
\ee

To conclude we have shown that any negative tension brane in an adS
background induces a bound state tachyon mode in the context of
Einstein-Gauss-Bonnet gravity.  It would be 
interesting to study if the instability portrayed here persists for
other spacetime geometries as well (see also \cite{us} where
Gauss-Bonnet gravity destabilises scalar field brane solutions).   
In string theory both Gauss-Bonnet corrections \cite{gross} 
and negative tension
boundaries, orientifold planes \cite{sagnotti}, \cite{polc} can appear. 
A typical example is that of
open type I $SO(32)$ string theory. Orientifolds there appear as
non-dynamical 8-branes of RR charge \cite{witt} and negative tension.
The Gauss-Bonnet term also
appears at the level of the disk since the underlying theory is S dual
to the $SO(32)$ heterotic string (where the Gauss-Bonnet term appears at
tree-level \cite{gross}).
Clearly the field content is richer and much more complex in this case, 
especially since the Gauss-Bonnet term would be coupled to a varying 
dilaton field.  Furthermore even if
the instability persisted as a bound state to the orientifold, then
the effective action approach could be unjustified. 
The stability issues for string theory backgrounds certainly
require further careful investigation.

A final issue is the endpoint of the instability.  
One can argue that the brane action may pick up induced gravity
terms (at the price of adding 2 more coupling constants to the
original action) and in this case we showed that one may avoid the
instability by restricting the according couplings (\ref{range}). 
Alternatively the RS I 
solution is not a stable vacuum of the action 
(\ref{Sbulk}) and therefore one 
may question what the true vacuum of such a theory is. 
For instance had we considered 4 dimensional 
dS or adS branes would the spin 2 instability persist ? 
%Note that even if spin 2
%fluctuations are stable for a de-Sitter domain wall, the radion will be
%unstable, \cite{bin} (see also \cite{sikivie}) in the sense 
%that the two branes will
%repel each other destroying any hierarchy argument.
Such a question on the stable vacuum may be interesting if stringent
constraints are required for the effective 4 dimensional cosmological
constant. This is a question we hope to be addressing in the near future.

We are particularly grateful to Pierre Binetruy and Jihad Mourad 
for their encouragement and enlightening discussions. We also thank 
Stephen Davis and Valery Rubakov  
for discussions at the early stages of this work. CC
thanks Luigi Pilo for discussions on mixed boundary conditions.


\begin{thebibliography}{99}

\bibitem{sagnotti}
%\cite{Sagnotti:1987tw}
A.~Sagnotti,
%{\em Open Strings And Their Symmetry Groups,}
%Talk presented at the Cargese Summer Institute 
%on Non-Perturbative Methods in Field Theory, Cargese, France, Jul
%16-30, 1987. 
published in Cargese Summer Inst.1987:0521-528 (QC174.45:N2:1987), 
arXiv:hep-th/0208020.
%%CITATION = HEP-TH 0208020;%%
%\cite{Polchinski:1987tu}
J.~Polchinski and Y.~Cai,
%{\em Consistency Of Open Superstring Theories},
Nucl.\ Phys.\ B {\bf 296}, 91 (1988).
%%CITATION = NUPHA,B296,91;%%


\bibitem{polc}
J.~Polchinski,
{\em String Theory, Vol. I and II}.
J.~Polchinski,
%{\em Dirichlet-Branes and Ramond-Ramond Charges,}
Phys.\ Rev.\ Lett.\  {\bf 75}, 4724 (1995).
%[arXiv:hep-th/9510017].
%%CITATION = HEP-TH 9510017;%%


\bibitem{RS1}
L.~Randall and R.~Sundrum,
%{\em A large mass hierarchy from a small extra dimension},
Phys.\ Rev.\ Lett.\  {\bf 83}, 3370 (1999).
%[arXiv:hep-ph/9905221].

\bibitem{GW}
E.g.: W.~D.~Goldberger and M.~B.~Wise,
%{\em Modulus stabilization with bulk fields},
Phys.\ Rev.\ Lett.\  {\bf 83}, 4922 (1999).
%[arXiv:hep-ph/9907447].
%\cite{Martin:2003yh}
%\bibitem{Martin:2003yh}
J.~Martin, G.~N.~Felder, A.~V.~Frolov, M.~Peloso and L.~Kofman,
%{\em Braneworld dynamics with the BraneCode}
arXiv:hep-th/0309001.
%%CITATION = HEP-TH 0309001;%%
%\cite{Lesgourgues:2003mi}
%\bibitem{Lesgourgues:2003mi}
J.~Lesgourgues and L.~Sorbo,
%{\em Goldberger-Wise variations: Stabilizing brane models with a bulk scalar}
arXiv:hep-th/0310007.
%%CITATION = HEP-TH 0310007;%%

\bibitem{Pol}
E.g.:
S.~B.~Giddings, S.~Kachru and J.~Polchinski,
%{\em Hierarchies from fluxes in string compactifications}
Phys.\ Rev.\ D {\bf 66}, 106006 (2002).
%[arXiv:hep-th/0105097].

\bibitem{gross}
D.~J.~Gross and J.~H.~Sloan,
%{\em The Quartic Effective Action For The Heterotic String},
Nucl.\ Phys.\ B {\bf 291} (1987) 41.
%%CITATION = NUPHA,B291,41;%%
%\bibitem{Mets}
R.~R.~Metsaev and A.~A.~Tseytlin,
%{\em Order Alpha-Prime (Two Loop) Equivalence Of The String Equations Of
%Motion And The Sigma Model Weyl Invariance Conditions: Dependence On The
%Dilaton And The Antisymmetric Tensor}, 
Nucl.\ Phys.\ B {\bf 293} (1987) 385.
%CITATION = NUPHA,B293,385;%%

\bibitem{Zwieb}
B.~Zwiebach,
%{\em Curvature Squared Terms And String Theories},
Phys.\ Lett.\ B {\bf 156} (1985) 315.
%%CITATION = PHLTA,B156,315;%%

\bibitem{Love}
D.~Lovelock,
%{\em The Einstein Tensor And Its Generalizations},
J.\ Math.\ Phys.\ {\bf 12} (1971) 498.
%%CITATION = JMAPA,12,498;%%

\bibitem{stephen}
S.~C.~Davis,
%{\em Generalised Israel junction conditions for a Gauss-Bonnet brane world,}
Phys.\ Rev.\ D {\bf 67}, 024030 (2003).
%[arXiv:hep-th/0208205].
%%CITATION = HEP-TH 0208205;%%\\
%\cite{Gravanis:2002wy}
%\bibitem{Gravanis:2002wy}
E.~Gravanis and S.~Willison,
%{\em Israel conditions for the Gauss-Bonnet theory and the Friedmann
%equation on the brane universe,}''
Phys.\ Lett.\ B {\bf 562}, 118 (2003).
%[arXiv:hep-th/0209076].
%%CITATION = HEP-TH 0209076;%%

\bibitem{radion}
C.~Charmousis, R.~Gregory and V.~A.~Rubakov,
%``Wave function of the radion in a brane world,''
Phys.\ Rev.\ D {\bf 62}, 067505 (2000).
%[arXiv:hep-th/9912160].

\bibitem{neu}
Y.~M.~Cho, I.~P.~Neupane and P.~S.~Wesson,
%{\em No ghost state of Gauss-Bonnet interaction in warped background},
Nucl.\ Phys.\ B {\bf 621} (2002) 388. 
%[hep-th/0104227]. 
K.~A.~Meissner and M.~Olechowski,
%{\em Brane localization of gravity in higher derivative theory},
Phys.\ Rev.\ D {\bf 65} (2002) 064017. 
%[hep-th/0106203].

\bibitem{us}
C.~Charmousis, S.~C.~Davis and J.~F.~Dufaux,
%``Scalar brane backgrounds in higher order curvature gravity,''
JHEP {\bf 0312}, 029 (2003).
%arXiv:hep-th/0309083.
%
%\cite{Zanelli:2002qm}
%\bibitem{chile}
%J.~Zanelli,
%{\em (Super)-gravities beyond 4 dimensions,}
%arXiv:hep-th/0206169.
%%CITATION = HEP-TH 0206169;%%

\bibitem{RS2}
L.~Randall and R.~Sundrum,
%{\em An alternative to compactification},
Phys.\ Rev.\ Lett.\ {\bf 83} (1999) 4690. 
%[hep-th/9906064].

\bibitem{LR}
J.~Lykken and L.~Randall,
%``The shape of gravity,''
JHEP {\bf 0006}, 014 (2000).
%[arXiv:hep-th/9908076].

\bibitem{dav}
H.~Davoudiasl, J.~L.~Hewett and T.~G.~Rizzo,
%``Brane localized curvature for warped gravitons,''
JHEP {\bf 0308}, 034 (2003)
%[arXiv:hep-ph/0305086].
%\bibitem{libanov}
S.~L.~Dubovsky and M.~V.~Libanov,
%``On brane-induced gravity in warped backgrounds,''
JHEP {\bf 0311}, 038 (2003).
%arXiv:hep-th/0309131.
%
%\bibitem{roy}
G.~Kofinas, R.~Maartens and E.~Papantonopoulos,
%``Brane cosmology with curvature corrections,''
JHEP {\bf 0310}, 066 (2003).
%arXiv:hep-th/0307138.

\bibitem{csaki}
C.~Csaki, M.~L.~Graesser and G.~D.~Kribs,
%``Radion dynamics and electroweak physics,''
Phys.\ Rev.\ D {\bf 63}, 065002 (2001)
%[arXiv:hep-th/0008151].

\bibitem{pilo}
L.~Pilo, R.~Rattazzi and A.~Zaffaroni,
%{\em The fate of the radion in models with metastable graviton},
JHEP {\bf 0007} (2000) 056
%[hep-th/0004028].

%\cite{Shtanov:2003um}
\bibitem{yuri}
Y.~Shtanov and A.~Viznyuk,
%``Linearized gravity on the Randall-Sundrum two-brane background with
%curvature terms in the action for the branes,''
arXiv:hep-th/0312261.
%%CITATION = HEP-TH 0312261;%%


%\cite{Bostock:2003cv}
%\bibitem{ruth}
%P.~Bostock, R.~Gregory, I.~Navarro and J.~Santiago,
%{\em Einstein gravity on the codimension 2 brane?,}
%arXiv:hep-th/0311074.
%%CITATION = HEP-TH 0311074;%%


%\cite{Polchinski:1995df}
\bibitem{witt}
J.~Polchinski and E.~Witten,
%{\em Evidence for Heterotic - Type I String Duality,}
Nucl.\ Phys.\ B {\bf 460}, 525 (1996).
%[arXiv:hep-th/9510169].
%%CITATION = HEP-TH 9510169;%%


%\cite{Kawai:1998ab}
%\bibitem{Kawai}
%S.~Kawai, M.~a.~Sakagami and J.~Soda,
%{\em Instability of 1-loop superstring cosmology,}
%Phys.\ Lett.\ B {\bf 437}, 284 (1998)
%[arXiv:gr-qc/9802033].
%%CITATION = GR-QC 9802033;%%


%\cite{Bachas:1999um}
%\bibitem{bachas}
%C.~P.~Bachas, P.~Bain and M.~B.~Green,
%{\em Curvature terms in D-brane actions and their M-theory origin,}
%JHEP {\bf 9905}, 011 (1999).
%[arXiv:hep-th/9903210].
%%CITATION = HEP-TH 9903210;%%


%\cite{Binetruy:2001tc}
%\bibitem{bin}
%P.~Binetruy, C.~Deffayet and D.~Langlois,
%{\em The radion in brane cosmology},
%Nucl.\ Phys.\ B {\bf 615}, 219 (2001).
%[arXiv:hep-th/0101234].
%%CITATION = HEP-TH 0101234;%%\\
%\cite{Gen:2000nu}
%\bibitem{Gen:2000nu}
%U.~Gen and M.~Sasaki,
%{\em Radion on the de Sitter brane,}
%Prog.\ Theor.\ Phys.\  {\bf 105}, 591 (2001).
%[arXiv:gr-qc/0011078].
%%CITATION = GR-QC 0011078;%%\\
%\cite{Frolov:2003yi}
%\bibitem{Frolov:2003yi}
%A.~V.~Frolov and L.~Kofman,
%{\em Can inflating braneworlds be stabilized,}
%Phys.\ Rev.\ D {\bf 69}, 044021 (2004).
%arXiv:hep-th/0309002.
%%CITATION = HEP-TH 0309002;%%

%\cite{Ipser:1983db}
%\bibitem{sikivie}
%J.~Ipser and P.~Sikivie,
%{\em The Gravitationally Repulsive Domain Wall,}
%Phys.\ Rev.\ D {\bf 30}, 712 (1984).
%%CITATION = PHRVA,D30,712;%%




\end{thebibliography}
\end{document}